\documentclass[aps,prb,superscriptaddress,reprint]{revtex4-1}
\usepackage{amsmath,amssymb} 
\usepackage{amsmath} 
\usepackage{amsfonts} 
\usepackage{bm} 
\usepackage[colorlinks=true,citecolor=blue,linkcolor=blue,filecolor=blue,urlcolor=blue]{hyperref} 
\usepackage[pdftex]{graphicx} 
\usepackage{braket} 
\usepackage{physics}
\usepackage{color}
\usepackage{comment}
\usepackage{mathtools}

\begin{document}
\title{Diffusive dynamics of fractionalized particles 
and the enhanced conductivity at the border of the neutral-ionic transition}
\author{Yuta Sakai}
\author{Chisa Hotta}
\affiliation{The University of Tokyo, Komaba, Meguro-ku, Tokyo 153-8902, Japan}
\begin{abstract}
We study the diffusive dynamics of the classical one-dimensional lattice model of 
mobile particles featuring the incoherent metallic state of the organic TTF-CA 
found in the vicinity of the neutral-ionic(NI)-transition. 
The particles are strongly correlated and feel the alternating site potentials, 
exhibiting uniform ionic (I) Mott insulating and neutral band insulating (N) phases 
when the Coulomb interaction and site potentials are large, respectively. 
We focus on the neutral-ionic domain walls (NIDW) activated by their competition, which is typically 
regarded as fractionalized particles. 
The finite temperature phase diagram reveals a thermodynamically stable NIDW phase
not found in previous literature that emerges at the triple point, where 
the transition line splits into two first-order lines toward the critical endpoints. 
We analyze the long-time behavior of dynamics of the NIDWs and find that it captures the diffusive transport of the system. 
The conductivity derived from the diffusion constant and the number population of NIDWs 
shows a strong enhancement inside the NIDW phase, 
which can explain the large conductivity at the NI crossover found in TTF-CA. 
\end{abstract}
\maketitle
\section{Introduction}\label{introduction}
Enhanced correlations near the boundaries between different material phases 
have provided a compelling and fruitful platform for condensed matter research. 
Intriguing aspects of such phases are often captured by the anomalous behavior of transport properties. 
The anomalous electrical resistivity linear-in-temperature resistivity beyond the Mott-Ioffe-Regel (MIR) 
limit known as ``bad metals" \cite{Bruin2013} are the longstanding topic across 
high $T_c$ cuprates\cite{Kordyuk2015,Keimer2015}, vanadium dioxide\cite{Allen1993,Qazilbash2006}, 
and alkali-doped fullerides\cite{Gunnarsson2003}. 
Similar behavior is found at much lower temperatures in heavy fermionic systems\cite{Trovarelli2000}
referred to as strange metals due to their non-Fermi liquid nature\cite{Schofield1999}. 
Perovskite and Ruddlesden-Popper manganites show a large sensitivity in response 
to an applied magnetic field\cite{Salamon2001}. 
Their electrical resistivity grows by orders of magnitudes as a consequence of the interplay of charge, 
spin and orbitals accompanying phase segregation or inhomogeneity\cite{Dagotto2001}. 
These phenomena are more or less governed by many integrants, 
which makes it overwhelming to establish a simple understanding of their underlying mechanisms beyond 
the phenomenology\cite{Hartnoll2014}.  
\par
To this end, organic materials can be an ideal platform, because they show as rich a phase diagram 
as the aforementioned materials while they are mostly based on the single molecular orbital picture 
and relatively strong correlation effects\cite{chisa2004}. 
Aside from the most famous $\kappa$-ET families that show Mott insulator, superconductivity, criticality, and spin liquid behaviors, 
TTF-CA has also captured a long-standing interest in featuring neutral-ionic (NI) transition, 
which embodies the competition of band insulator and Mott insulator\cite{Sunami2022}. 
The present paper focuses on the anomalous enhancement of electrical conductivity observed in TTF-CA 
at the NI crossover region that develops on the higher temperature part of the phase boundary\cite{Takehara2019}. 
\par
Experimentally, the anomalous conduction of TTF-CA has been attributed to topological low-energy excitation 
called neutral-ionic domain wall (NIDW). 
The NIDWs carry fractionalized elementary charges, $\pm e/2$, 
defined as the averaged charge density of neighboring two sites on both sides of the bond. 
In an applied dc electric field, thermally excited NIDW pairs carry fractional charges of opposite signs. 
In the most elementary picture, they may drift in opposite directions and contribute to the electrical conductivity. 
Such fractionalization is typical of one- or quasi-one-dimensional systems, 
e.g. XXZ spin-1/2 model with a finite spin gap\cite{Mayr2006} and both the bosonic and fermionic models 
on the anisotropic triangular lattice\cite{Kohno2007,Hotta2008}, 
which are sometimes called fractons, where they typically yield a continuum in their excitation spectrum. 
For TTF-CA, it was discussed a long time ago\cite{Nagaosa1986-2} using the XXZ spin-1/2 model as an effective model 
in the limit of strong on-site Coulomb interactions and potentials. 
\par
Indeed, the studies on TTF-CA go back to the 1980's and the experimental phase diagram for temperature 
and pressure has been clarified, mostly based on transport\cite{Takehara2019}, 
NQR\cite{Takehara2018}, NMR\cite{Sunami2018}, and optical measurements\cite{Masino2007,Okamoto1989}. 
At ambient pressure upon cooling, the neural phase transforms to the ionic state at 81 K to gain the Madelung energy 
including the long-range Coulomb interactions. 
By applying pressure, the high-temperature neutral phase crossovers to the paraelectric ionic state where 
the charge transfer occurs from TTF to CA, 
and at lower temperatures, the lattice distortion associated with the spin degrees of freedom of 
the Mott state drives the system to a ferroelectric ionic phase by breaking the inversion symmetry. 
\par
In theories, the ionic extended Hubbard model is applied, which includes 
the on-site and nearest neighbor Coulomb interactions $U$ and $V$, the uniform transfer integral $t_{r}$, 
and the alternating difference in the energy level $\Delta$ between TTF and CA \cite{Nagaosa1986-2}. 
The basic feature of this model was studied in the initial series of works based on the quantum Monte Carlo simulation (QMC), 
where the degree of charge transfer denoted as $\rho$ shows a discontinuity at the phase transition at low temperatures. 
For the ionic Hubbard model without $V$, 
a more extensive studies have followed that clarified the ground state phase diagram on the plane of $U$ and $\Delta$, 
where the competition happens between the band insulator and the Mott insulator, 
namely, neutral and ionic phases, at around $\Delta \sim U$. 
Fabrizio, {\it et.al.} proposed a scenario based on field theory that there is a spontaneously dimerized phase 
in between the band insulator and the Mott insulator\cite{Fabrizio1999}. 
The dimerization destroys the spin-density-wave (SDW) Mott state and transforms it into the bond charge-density-wave (BCDW) state 
before entering the CDW band insulator. 
Despite several controversial numerical results in early periods\cite{Takada2001, Lou2003,Anusooya-Pati2001,Wilkens2001,Gidopoulos2000} 
later numerical and analytical works have confirmed the existence of very narrow dimerized phase\cite{Torio2001,Zhang2003,Manmana2004,Tsuchiizu2004}. 
This means that without the introduction of lattice dimerization, there is an underlying tendency to 
form a dimer toward the magnetically gapped singlet state with ferroelectricity\cite{Wilkens2001}. 
The latest QMC study at finite temperature\cite{Otsuka2012} implementing the lattice dimerization showed 
the double free-energy-minima in the low-temperature dimerized ionic phase. 
There, the dimerization disappears at high temperatures after they may undergo a first-order transition 
with a jump in the degree of charge transfer and hysteresis 
\footnote{The role of dimerization on the order of transition is not clear in the literature. 
The ionic-to-neutral transition can be of first order when $V$ is present, or even by the choice of the boundary conditions at finite size. }. 
The strong coupling limit of the ionic Hubbard model was studied by the effective spin-1 model
\cite{Legeza2006} in agreement with a narrow intermediate phase\cite{Tincani2009}. 
\par
The model is also studied in higher spatial dimensions, 
where the BCDW mentioned above  in the ground state phase diagram disappears, 
and instead, the metallic phase appears in between the band and Mott insulating phases\cite{Bouadim2007,Paris2007}. 
The metallic state is stabilized by $\Delta$ in two dimensions\cite{Bouadim2007,Paris2007}, 
while in infinite dimensions the dynamical mean-field theory shows the opposite\cite{Bag2021}. 
The dc conductivity obtained by the fermionic QMC\cite{Paris2007} exhibits a crossover 
from a metallic to the insulating state as the temperature is lowered, 
while in overall, the theory is limited to the temperature range of $T\lesssim 0.5t_{r}$. 
Other extensions of models including the doping of carriers or modification of band structures 
reports further enriched ground state phases\cite{Garg2014,Bag2021}, 
indicating a large room to study the dynamics in the NI systems. 
\par
Recently, some dynamic aspects of the NI transition was clarified by the experimental works on TTF-CA 
at the two characteristic phases in the temperature range of $T\sim 280-300$ K. 
Based on the infrared spectroscopy \cite{Okamoto1989}, NQR \cite{Takehara2019} and NMR \cite{Sunami2018} experiments, 
it is argued that, in the dimerized paraelectric ionic phase at $P\sim 10-35$ kbar, 
there arise dynamical ferroelectric domains carrying opposite polarizations with their neighboring domains 
separated by the defects in the dimerization periods, which they call spin solitons. 
At the crossover region between the neutral and paraelectric ionic state, $P\sim 8-10$ kbar, 
a large enhancement of conductivity whose magnitude is comparable to the metallic one is observed\cite{Takehara2019}, 
which is attributed to NIDW conductivity. 
In theories, there had been some works based on the phase Hamiltonian and bosonization techniques 
\cite{Fukuyama2016,Tsuchiizu2016} 
to explain what kind of topological excitations and the gaps are available. 
However, the dynamic aspect of the NI system has not yet been explored so far, 
particularly in relevance with the dynamical inhomogeneity due to NIDWs and their many-body effects. 
\par
In the present paper, we study the one-dimensional classical model in the strong coupling limit of the ionic extended Hubbard model. 
We apply a Monte Carlo (MC) based Glauber dynamics 
and evaluate the dissipative nature of the NIDWs, 
showing that the NIDWs thermally activated in pairs across the charge transfer gap 
can fluctuate and propagate for a substantial timescale until they meet another NIDW carrying an opposite charge and disappear in pairs. 
The basic treatment given here is to find numerically in the classical limit of the quantum microscopic model 
that the hydrodynamic equations to describe a coarse-grained thermodynamic version of the microscopic model actually hold. 
We extract the electrical conductivity using the Einstein relation, 
finding that it is indeed enhanced at the phase boundary 
due to the substantial increase of the population of NIDWs, 
despite that the diffusion constant is suppressed by the correlation effect. 
Our calculation proves that such NIDW conductivity can happen without a lattice dimerization or quantum fluctuation effects, 
and that there is a thermodynamically stable NIDW phase at a relatively high temperature at least in the classical limit. 

\begin{figure}[tbp]
  \includegraphics[width=8.5cm]{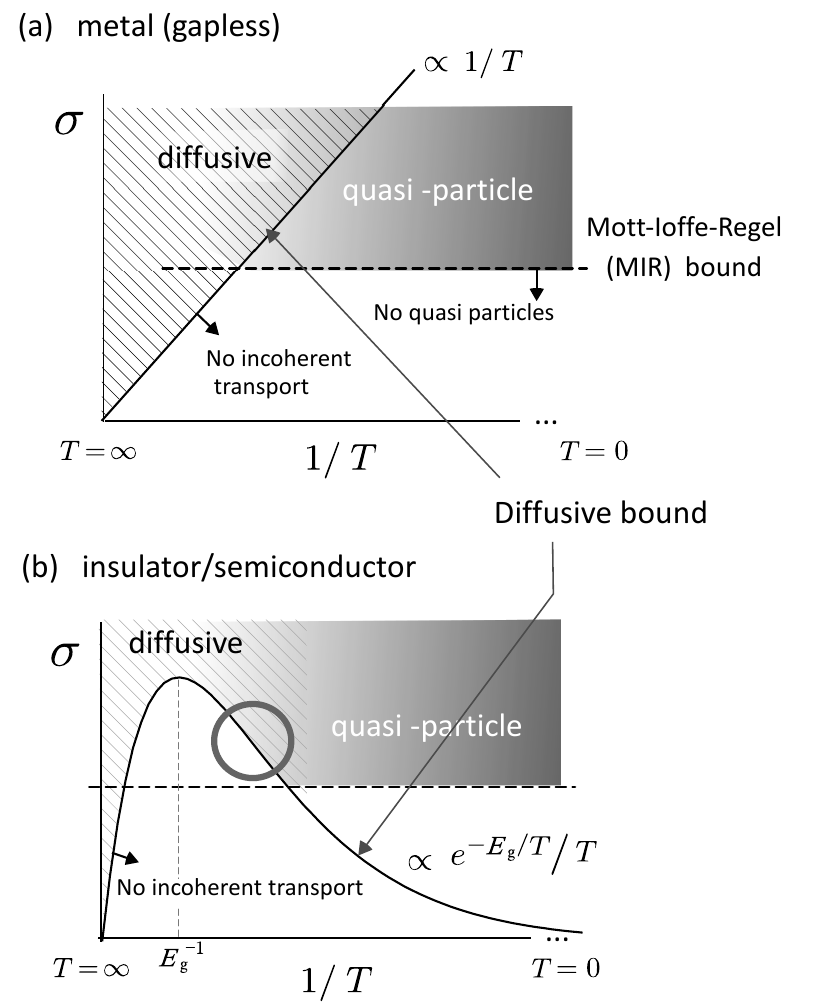}
  \caption{Schematic illustration of the characteristics of transport 
  in the conductivity versus inverse-temperature plane 
  for (a) the metallic (gapless) and (b) the insulating/semiconducting (with a gap $E_g$) cases. 
  Broken line is the MIR bound, below which we no longer find a quasi-particle, 
  and the solid line gives the diffusive bound below which we no longer expect the incoherent transport. 
  Open circle marks the location of the present system.}
  \label{cartoon}
\end{figure}
\section{Dynamics at finite temperature}\label{Sec:dynamics}
Before entering, we briefly overview the characteristics of the dynamics 
in quantum many-body systems and how they are treated in the existing framework. 
\par
Figure~\ref{cartoon} schematically illustrates the nature of dynamics that varies with temperature 
for different conductivity range. 
The MIR line gives the lower bound of conductivity, $\sigma$, 
below which the quasi-particles can no longer survive as the mean-free path of the conducting quasi-particles 
becomes shorter than the de Broglie wavelength. 
Recently, another bound for the incoherent transport is discussed in Ref.[\onlinecite{Hartnoll2014}], 
given by $\sigma \propto 1/T$ (see Fig.~\ref{cartoon}(a)), 
phenomenologically explaining the fact that the incoherent metals approximately saturating 
the bound has a linear-in-temperature electrical resistivity. 
This behavior continues even at low temperatures in bad metals above the MIR bound. 
In our system, a large energy gap $E_g$ is present, 
and the diffusive bound shall be modified to $\sigma \propto e^{-E_g/T}/T$ 
as shown in Fig.~\ref{cartoon}(b), because the carriers are activated over a gap 
whose density is given as $n \propto e^{-E_g/k_BT}$. 
We consider the region marked in circle; $T\lesssim E_g$ and above the MIR bound,  
where the quasi-particles are still well-defined but exhibits a diffusive nature of transport. 
\par
It is known that the dynamics of the quantum many-body systems are generally elusive, even in one dimension (1D). 
When we start from the classical product state and quench the system 
e.g. setting one spin as down and the other as all up in the quantum spin-1/2 system, 
and track the time($t$)-dependence, 
the correlation range spreads linear-in-$t$ in a short timescale inside the Lieb-Robinson bound\cite{Lieb1972} as demonstrated experimentally in cold atoms\cite{Jurcevic2014}. 
In such a case, the entanglement entropy follows an area-law bounded by $t$\cite{Eisert2006}, 
which means that the numerical resource required for the calculation blows up with time. 
Indeed, the quantum many-body calculations can typically follow the exact dynamics only up to $t < 1$ in a unit of inverse the interaction strength\cite{Kaneko2022}. 
After a long time, most of the system thermalizes to an equilibrium state that follows an entanglement volume law, 
which is usually beyond the numerical description. 
The difficulty is partially overcome by the framework using the thermal pure state, 
where we were able to track almost numerically exactly such 
dynamics up to $t\sim 100$\cite{Endo2018}. 
However, the method is limited to a system size of $N\lesssim 30$. 
\par  
Still, there are several indications that the quantum many-body dynamics can be treated reasonably 
within a classical framework. 
In the transverse Ising model, 
the semiclassical description of quantum relaxation dynamics assuming 
the ballistically moving classical quasi-particles\cite{Sachdev1997} 
agrees surprisingly well with the exactly solved time-dependent processes \cite{Rieger2011}. 
The Glauber dynamics extended to quantum Monte Carlo simulation successfully extracts 
the dynamical exponent at low temperatures near the phase transition\cite{Hotta2023}. 
In nonintegrable models, quasi-particles encountering few collisions follow the Boltzmann equation, 
which describes well the local equilibration processes. 
These results indicate that the low-temperature quantum dynamics at relatively short timescale is replaced by that of the classical ones by properly assuming the carriers and 
the relaxation mechanism. 
\par
On the other hand, the late-time transport  
is known to be governed by classical hydrodynamics which respects the
structure of the conservation laws of the system\cite{Mukerjee2006}. 
When the system reaches a global equilibrium, 
the behavior of quasi-particles is predicted well 
by the hydrodynamics of the particle density following the stochastic linearized diffusion equations 
with some higher order correction terms if necessary\cite{Lux2014,Kawano2024}. 
These hydrodynamics build up a long timescale and contribute to the Drude component. 
The conservation laws work in such a way that for 1D quantum spin systems, 
The ac and dc conductivities at high temperatures are determined solely 
by the dynamical exponent $z$\cite{Dupont2020}. 
The quasi-particles spread with time $t$ as $|x|\sim t^{1/z}$, 
and the $z=2$ case corresponds to the normal diffusive transport, 
$z=1$ and $3/2$ are the ballistic and superdiffusive ones, respectively. 
\par
Based on these considerations, we map out our system in Fig.~\ref{cartoon}(b) 
where the classical hydrodynamics of our quasi-particle, a NIDW, 
is governed by the diffusion constant and the carrier densities. 
The diffusive $z=2$ behavior of NIDWs is indeed found in our results. 
\section{Model and methods}\label{model_methods}
\subsection{Model}
\begin{figure}[tbp]
  \includegraphics[width=8.5cm]{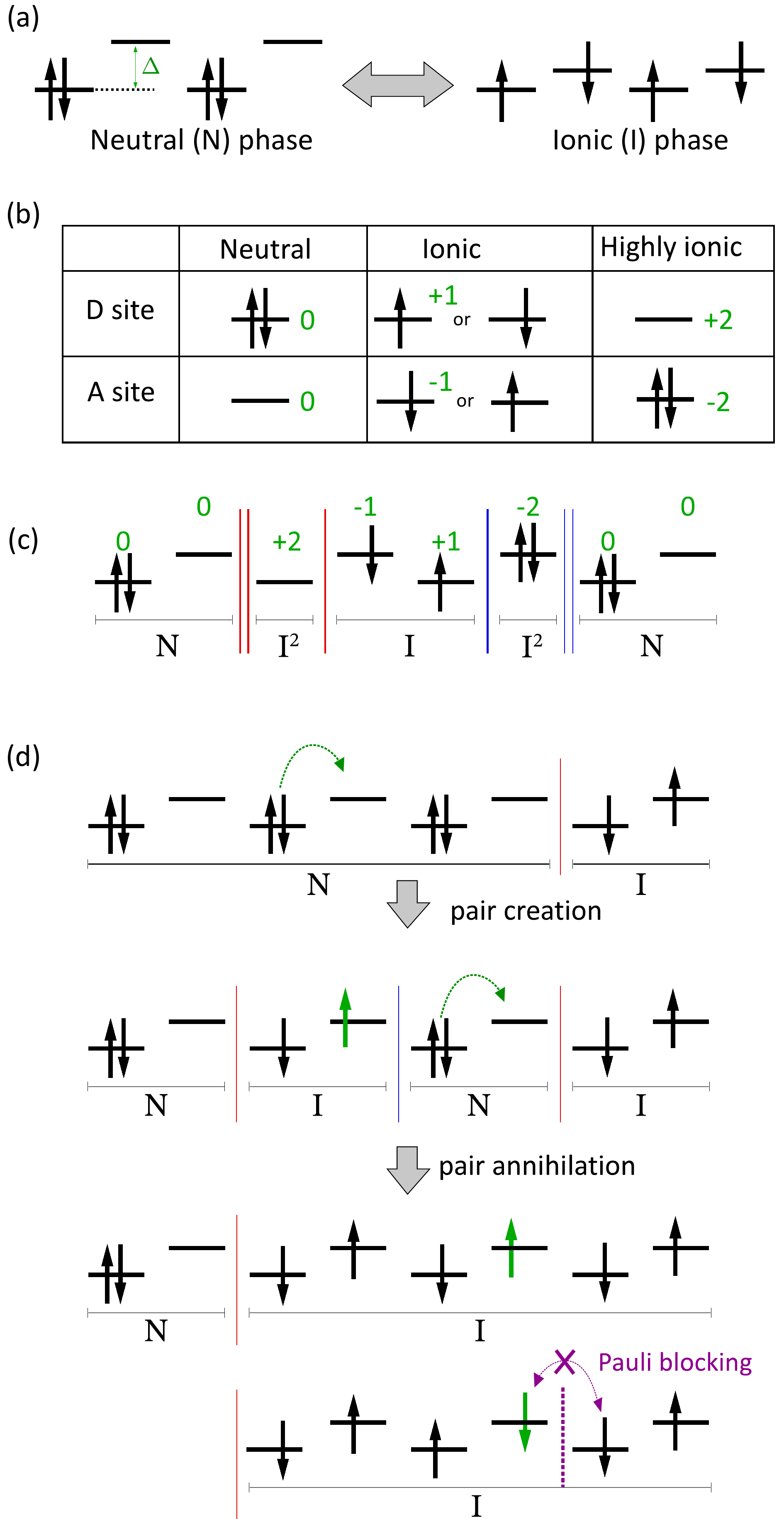}
  \caption{(a) Schematic illustration of NI transition.
  (b) Particle configurations on D and A sites, whose charge densities are measured from a neutral(N) state 
  with two particles on D sites as being zero. 
  (c) Schematic illustration of the distribution of particles with NIDWs, shown in vertical bars 
   which carry $\pm 1/2$ (red/blue) charges. 
  (d) Example of creating or annihilating the NIDWs in pairs. 
}
  \label{NIDW}
\end{figure}

We consider a model of classical particles carrying up and down spins in 
1D chain consisting of two sublattices denoted as A and D, 
whose Hamiltonian is given as 
\begin{align}\label{eq:HDA}
  \mathcal{H}_{\rm cl}= U\sum_{i=1}^{N}n_{i,\uparrow}n_{i,\downarrow} + V\sum_{i=1}^{N}n_{i}n_{i+1} 
+\Delta \sum_{j\in \text{A}} n_{j}, 
\end{align}
where $n_{i\sigma}=0$ or $1$ is the number of particles with $\sigma=\uparrow,\downarrow$ spins on-site $i$ 
interacting via on-site and nearest neighbor Coulomb interactions, $U$ and $V$. 
We set the energy level of site A to be higher than that of site D by $\Delta$, 
as shown schematically in Fig.~\ref{NIDW}(a). 
We set $U=7.5$ and $V=3.5$ throughout this paper following the previous works on TTF-CA\cite{Nagaosa1986-2}. 
\par
Previous studies on the NI transition have mainly focused on the ionic extended Hubbard model given as 
$\mathcal{H}_{\mathrm{IEHM}}=\mathcal{H}_{\rm cl}+\mathcal{H}_{\rm kin}$ 
which, in addition to Eq.(\ref{eq:HDA}), includes the quantum kinetic hopping term given as
${\mathcal H}_{\rm kin}= -t_r\sum_{i=1}^{N}\sum_{\sigma=\uparrow,\downarrow}(c_{i,\sigma}^{\dagger}c_{i+1,\sigma}+\text{h.c.})$ 
with the electron creation/annihilation operators, $c_{i,\sigma}^{\dagger}/c_{i,\sigma}$, 
where the aforementioned values of $U$ and $V$ are taken as 
a unit of $t_r=1$ whose values are assumed as 0.1-0.2eV\cite{Ishibashi2010}. 
Our model serves as its strong coupling limit $t_r/U \rightarrow 0$, 
and accordingly, we assume that the particles follow Pauli's principle. 
The NI transition is characterized by the degree of charge transfer from site D to site A given as, 
\begin{equation}
\rho= 2\sum_{j\in \text{A}} n_{j}/N, 
\end{equation}
which is called ionicity. 
Although $\rho$ can range between 0 and 2 in principle, $\rho>1$ is unlikely, 
because the two limiting cases, the perfect N and I phases, 
have $\rho=0$ and 1, respectively.  
We consider $\mathcal{H}_{\rm cl}$ at half-filling, namely the total number of particles is $N_e=N$ 
consisting of the same number of particles with up and down spins, 
and apply a periodic boundary condition. 
The NI transition is understood as a consequence of the competition between $U$ and $\Delta$. 

\subsection{Classical Monte Carlo dynamics}
\label{MC_dynamics}
We are concerned with the dynamical properties of $\mathcal{H}_{\rm cl}$, 
which can be captured within the framework of Glauber dynamics of the classical Monte Carlo(MC) calculation\cite{Glauber1963}. 
Glauber dynamics was first introduced to describe the time evolution of the Ising model by regarding the local spin flip as ``time" step, 
which constitutes a stochastic continuous Markov chain. 
This type of dynamic interpretation of MC sampling is generally known to be valid for the system 
which has no system-encoded intrinsic rule dominating the time evolution of the concerned degrees of freedom, 
instead, which can be modeled to evolve with stochastic dynamics 
induced by a weak coupling to a heat bath\cite{Binder1992,Binder1997,Binder2022}. 
Here, we call it (classical) MC dynamics. 
The MC dynamics has been applied to various classical systems so far, such as a Brownian motion of polymers\cite{Baumgartner1983}, 
relaxation phenomena in quadrupolar glasses\cite{Carmesin1987}, and a diffusion process of interstitial atoms in an alloy\cite{Kehr1981,Kehr1989} 
(for more information, see Refs.[\onlinecite{Binder1992}] and [\onlinecite{Binder1997}]). 
The aforementioned works using dynamic MC simulations aim to calculate dynamic properties as a function of Monte Carlo timestep (MCS) 
such as autocorrelation functions or a mean square displacement (MSD) of particles, 
and by analyzing those properties, a characteristic timescale and a diffusion constant are obtained. 
Their results show that MC dynamics can safely capture the dynamical properties of 
the systems that do not afford molecular dynamics (MD) simulations built on the equation of motion. 
Still, for some cases, the acceptance ratio at each MCS can be extremely low, which hinders the calculation, 
for which a similar framework called kinetic MC was developed\cite{Bortz1975}. 
We tested and found that our model has a high enough acceptance ratio to sustain the simpler Glauber dynamics. 
\par
We perform an MC calculation by the following steps; \\
(1) Randomly choose one particle with spin $\sigma$ and the direction to hop, 
\\
(2) If there already exists a particle with the same spin in the destination, discard the trial. 
Otherwise, transfer the particle with the transition probability of the heat-bath method, 
\begin{equation}\label{eq:transition_prob}
  P(\Delta E)=\frac{1}{1+e^{\Delta E/k_{B}T}}, 
\end{equation}
where $\Delta E$ is the energy difference between the states before and after the hop. \\
We repeat (1) and (2) over $N_e/2$ times per $\sigma=\uparrow,\downarrow$, 
which constitute a single MCS. 
We prepare an I-state as an initial state and perform the calculation over typically $N_{\rm MCS}\sim 10^9$ steps, 
while discarding the first $10^8$ MCSs in calculating the phase diagram. 
For the dynamical calculations, we set 
$(N_{\rm MCS},N)=(10^9,512)$ and $(5\times10^8,1024)$ for the calculation of $n_{\rm dw}, D$ and $\sigma_{\rm dw}$, 
and $N_{\rm MCS}=1\times10^9$ for both $N=512$ and $N=1024$ in obtaining $\sigma_e$, 
both of which the $10^6$ MCSs are discarded as a relaxation. 
We confirm that the results do not depend on the initial configurations. 
\subsubsection{Charge dynamics in an electric field}
We apply two different ways of analysis to evaluate the thermally activated electrical conductivity at finite temperatures. 
The first method is to introduce the effect of the electric field as a constant energy potential bias ${\cal E}$. 
This corresponds to adding $+e{\cal E}$ to $\Delta E$ in Eq.(\ref{eq:transition_prob}) 
when calculating the transition probability of hopping the particle to the right-hand side, and adding $-e{\cal E}$ for the left. 
The site potential for charge $-e$ is effectively higher on the right 
and serves as an electric field in the right direction 
that drives the particles toward the left. 
Similar simulations on classical transport using the MC dynamics with field gradient are implemented 
in the earlier works including Kawasaki dynamics\cite{Kawasaki1966}. 
In this framework, the current ${\cal I}$ can be defined as the number of particles passing through 
the periodic boundary divided by the total MCSs of the duration. 
We define the conductivity as 
\begin{equation}\label{eq:sigmae}
\sigma_e={\cal I}/{\cal E}
\end{equation} 
for a small enough electric field ${\cal E}$ 
that allows ${\cal I}/{\cal E}$ to be nearly independent of ${\cal E}$. 
\subsubsection{Diffusion of NIDW}
The second method is to estimate the conductivity of NIDW, denoted as $\sigma_{\rm dw}$, 
assuming that each NIDW carries a charge of $\pm e/2$. 
For this purpose, we calculate the population density of NIDW denoted 
as $n_{\rm dw}$ and the diffusion constant $D$. Here, the electric field is not applied. 
The following analysis is guaranteed when the NIDW shows diffusive dynamics, 
for which the Einstein relation holds. 
Our results indeed apply to this case. 
\par
To measure the charge density on each site in units of elementary charge $e$, 
we set the N-state as a charge-neutral vacuum state, 
i.e. with net zero charge for both D and A sites 
which are doubly occupied by particles and vacant, respectively. 
Figure~\ref{NIDW}(b) summarizes the realization of states and their charge densities.  
In the I state, each of the D and A sites is occupied by a single particle, 
which correspond to the charge density of $+1$ and $-1$, respectively, 
where the minus sign denotes the ``hole" measured from a vacuum.   
During the simulation, the D site can sometimes be vacant and the A site be doubly occupied 
which are denoted as $+2$ and $-2$, respectively. 
The ones with $\pm 2$ are called "highly ionic" (I$^2$) and because their energies are much higher than 
other configurations observed rarely but with a finite probability (see Fig.~\ref{NIDW}(c)). 
The charge density on each bond is given by the average of charges of both sides, 
which can take the values of $0$, $\pm 1/2$ and $\pm 1$. 
The amount of charges carried by the NIDWs is $\pm 1/2$. 
Accordingly, the bond that carries $\pm 1$ charge has two NIDWs. 
The NIDWs are created in pairs at some bond and are assigned $\pm 1/2$ charges as shown in Fig.~\ref{NIDW}(d). 
Once created, the NIDW randomly moves around 
until it meets another NIDW charged with the opposite sign and they disappear in pairs. 
By analyzing the trajectory of all the NIDWs created during the whole MC process, 
we can obtain $n_{\rm dw}$ and $D$ as we explain in the following. 
\par
Each NIDW has their own lifetime, and suppose that $\tau_{\rm ave}$ is the value averaged over 
all NIDWs that appear during the entire MCSs. 
The total number of NIDW generated is measured as ${N}_{\rm ave}^{\rm tot}$, 
regardless of the lifetime of NIDWs. 
Using these values, the NIDW density is defined as the average number density of existing NIDWs 
at a certain point in time as 
\begin{equation}\label{eq:number_density}
  n_{\rm dw} = {N}_{\rm ave}^{\rm tot}\frac{\tau_{\rm ave}}{N_{\rm MCS}} \frac{1}{N}. 
\end{equation}
It is natural to expect that the dynamics of NIDW follow the standard diffusion equation, 
in which case the following relationship applies;  
\begin{equation}\label{eq:D} 
  D = \lim_{t\rightarrow\infty} \frac{\langle x_{t}^{2} \rangle}{2t}. 
\end{equation}
Here, $x_t$ is the displacement of NIDW from the generated position in a unit of lattice constant 
after the elapsed time $t$, and $\langle x_{t}^{2} \rangle$ is the mean square displacement (MSD). 
The sample average $\langle \cdot \rangle$ is taken over all the NIDWs trajectories with the lifetime $\tau > t$. 
Once we obtain the diffusion constant, the mobility $\mu$ of NIDW is obtained 
by applying Einstein relation as 
\begin{equation}\label{eq:mu}
  \mu=\frac{eD}{2k_{B}T}. 
\end{equation}
Here, only for these equations we introduce explicitly the elementary charge $e=1$ for clarification, 
where we apply $e/2$ as the absolute effective charge of NIDW. 
Accordingly, the electrical conductivity $\sigma_{\rm dw}$ is calculated as 
\begin{equation}\label{eq:sigmadw}
 \sigma_{\rm dw} = n_{\rm dw} \frac{e}{2} \mu = \frac{ e^2 n_{\rm dw} D}{4k_{B}T}. 
\end{equation}
\begin{figure*}[tbp]
  \includegraphics[width=18cm]{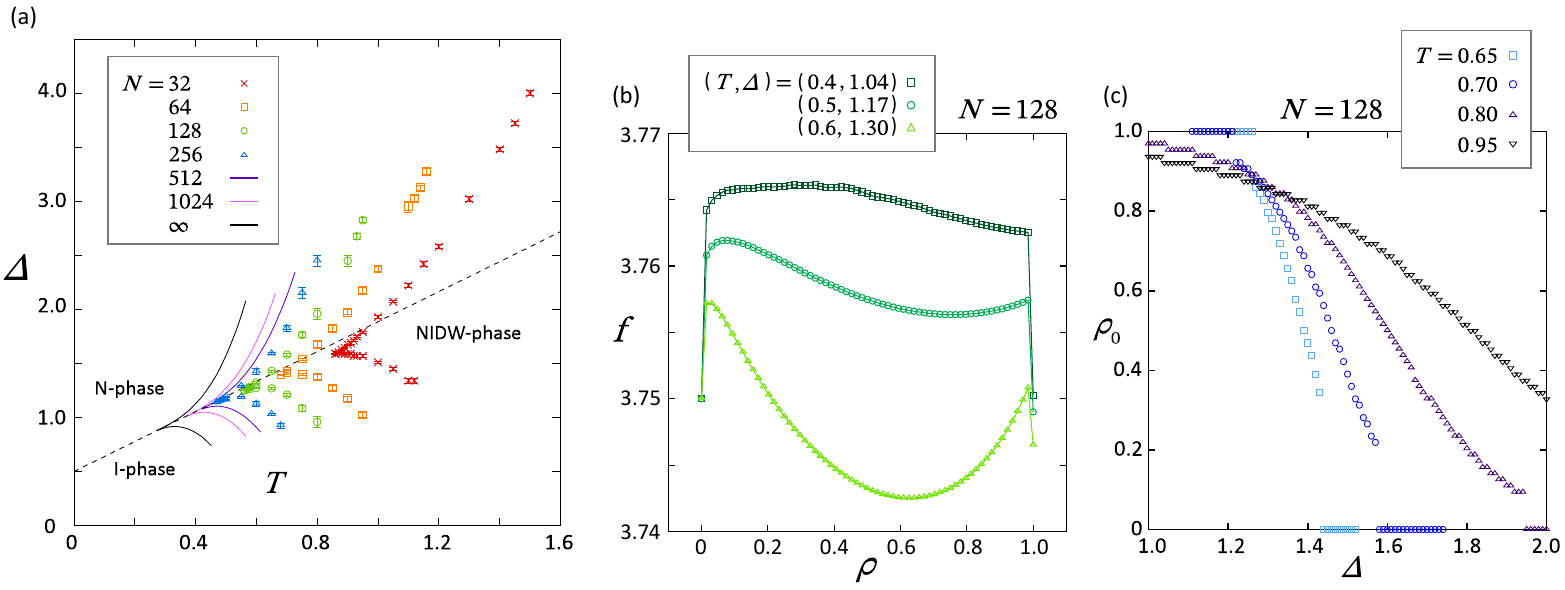}
  \caption{
(a) Finite temperature phase diagram of Eq.(\ref{eq:HDA}) at $U=7.5$ and $V=3.5$. 
(b) Free energy landscape $F(\rho;T,\Delta)$ at $N=128$ as a function of $\rho$ 
    for three different phases that have minimum point at $\rho_0=0,1$ and $0.625$. 
(c) The location of the free-energy minimum, $\rho_0$, 
    for $T=0.65, 0.70, 0.80$, and $0.95$ at $N=128$.
}
\label{phasediag}
\end{figure*}
\begin{figure}[tbp]
  \includegraphics[width=8cm]{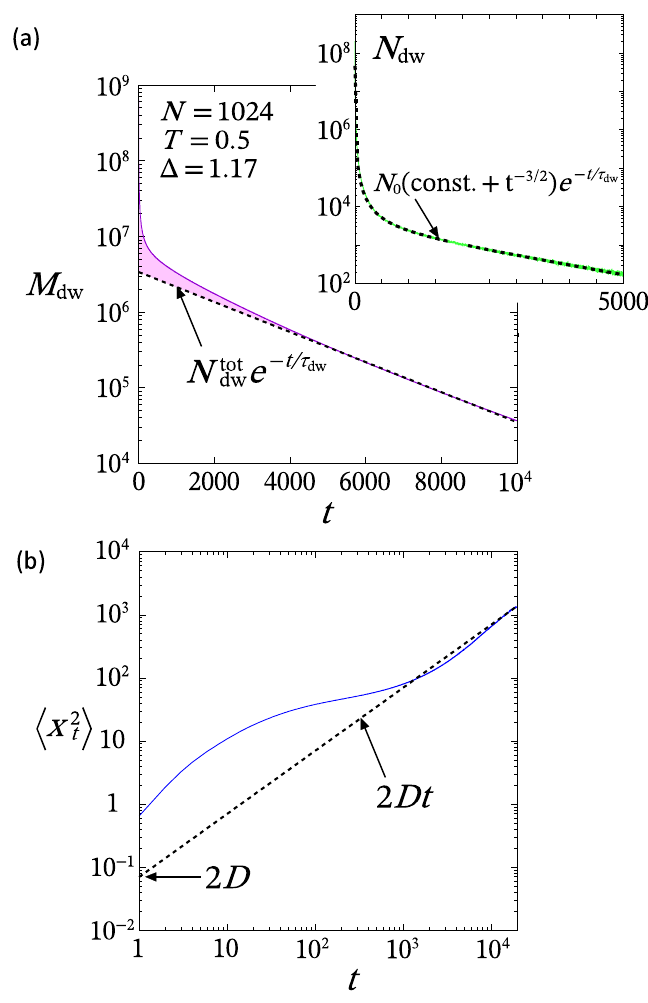}
  \caption{(a) $M_{\rm dw}$ in Eq.(\ref{cumulative_dist}) numerically obtained and averaged 
  as a function of $t$ (time step after NIDW was created) for $\Delta=1.17$, $T=0.5$ and $N=1024$. 
  The shaded region is the contribution from the short-lifetime NIDWs showing pair-recombination, 
  and the broken line is Eq.(\ref{eq:mdw-exp}) with $\tau_{\rm dw}=2.19\times 10^3$ and $N_{\rm dw}^{\rm tot}=3.39\times 10^6$.
  Inset shows the corresponding values of $N_{\rm dw}$, which is the lifetime distribution of NIDWs. 
  The broken line in the inset is the fitted curve using Eq.(\ref{eq:ndw-fit}). 
  (b) $\langle x_t^2\rangle$ as a function of $t$ for $\Delta=1.17$, $T=0.5$ and $N=1024$.
  The broken line is the result of the data fitting with Eq.(\ref{eq:D}) in the range of $t\ge1.5\times10^4$, where we find $D=3.5\times10^{-2}$}.
 \label{analysis}
\end{figure}
%
\section{Results}\label{results_discussions}
\subsection{Finite temperature phase diagram}\label{subsec:phase_diagram} 
We first present the phase diagram on the plane of $T$ and $\Delta$ in Fig.~\ref{phasediag}(a). 
At zero temperature, the NI transition takes place at $\Delta=U-2V$, 
which is determined by the difference in the ground state energies of the N and I phases. 
At finite temperature, the I phase is expected to be stabilized by the additional entropic term 
originating from the spin degeneracy of ${}_N \mathrm{C}_{\frac{N}{2}}$. 
Assuming that both I and N phases are homogeneous, free of any domains or defects, 
the phase boundary in the thermodynamic limit is obtained as 
\begin{equation}
\label{eq:boundary}
\Delta_c= \left( U-2V \right) + \left( 2\ln{2} \right)T_c, 
\end{equation} 
which is shown in the broken line in the phase diagram. 
\par
To unbiasedly elucidate the phase diagram, we perform an MC calculation and take the histogram 
of distribution of states against $\rho$ for given $T$ and $\Delta$, 
denoted as $H(\rho;T,\Delta)$. 
Because the distribution scales with the partition function, 
we can safely obtain the free energy profile as, 
$F(\rho;T,\Delta) = -T \ln H(\rho;T,\Delta) + C$, 
and then we can determine the value of the unknown constant $C$ using the value of 
free energy density of N phase, $f(\rho=0,T,\Delta)=U/2$. 
The free energy density for each value of $\rho$ is shown in Fig.~\ref{phasediag}(b). 
\par
At low temperatures, a two-minima structure of the free energy is observed. 
However, at relatively high temperatures near the NI phase boundary, 
three local minima appear, which means that another metastable phase appears 
and compete with the N and I phases. 
By analyzing these free energy landscapes, we extract the value $\rho=\rho_0$ that has the lowest value of 
$F(\rho;T,\Delta)$ as the thermal equilibrium state and obtain the phase diagram, 
where we denote the three phases as N phase ($\rho_0 =0$), I phase ($\rho_0 =1$) and NIDW phase ($0< \rho_0 <1$). 
The numerically obtained NI phase boundary at low temperatures almost coincides with the broken line, 
Eq.(\ref{eq:boundary}), within the error bars and are not shown for simplicity. 
\par
The phase boundaries between the NIDW-phase and the other two phases obtained by the MC calculations 
are shown for $N=32, 64, 128$, and $256$. 
We extrapolate these results by the polynomial fitting and obtain the boundaries for $N=512, 1024$, 
and the bulk limit $N\rightarrow \infty$ 
as a reference for the dynamical simulation we see shortly. 
We find in the free energy landscape that these transition lines are of first order, have two endpoints, 
and meet at the triple point reminiscent of the water. 
Figure~\ref{phasediag}(b) shows the example of three cases, $\rho_0=0,1$ and $0.625$, 
which correspond to the N, I, and NIDW phases, respectively. 
At high temperatures above the two endpoints, these phases show crossover behavior. 
\par
By extracting $\rho_0$ which gives the minimum of $f(\rho;T,\Delta)$, we obtain Fig.~\ref{phasediag}(c) 
as functions of $\Delta$ for several choices of $T$. 
When crossing the first order transition line, $\rho_0$ shows a clear discontinuity, 
which does not vanish with increasing system size, as we confirm by the extrapolation curves in the phase diagram. 
The values of jump decrease as the state approaches the endpoints and 
we find continuous changes with $\Delta$ to the I phase at $T=0.804$ and to I and N phases at $T=0.95$, 
which are the temperatures of the two endpoints for $N=128$.

\subsection{Diffusion of NIDWs}
\subsubsection{Two types of NIDWs}\label{results_nD}
In evaluating $n_{\rm dw}$, it is useful to know how observables related to each NIDWs behave 
before evaluating Eq.(\ref{eq:number_density}). 
For this purpose, we first consider the lifetime distribution of NIDWs denoted as $N_{\rm dw}(t)$, 
and the cumulative lifetime distribution is defined as 
\begin{equation}\label{cumulative_dist}
  M_{\rm dw}(t) = \sum_{t'=t+1}^{\infty} {N}_{\rm dw}(t'). 
\end{equation}
Here $t$ is the time step measured for each NIDW from its creation, where we take into account all of the NIDWs 
that appeared during the measurements over the entire MCSs, i.e. over $N_{\rm MCS}\sim 10^9$ steps. 
Figures~\ref{analysis}(a) and \ref{analysis}(b) show ${M}_{\rm dw}$, ${N}_{\rm dw}$ and $\langle x^2\rangle$ as functions of $t$. 
Some of the NIDWs recombine with the counterpart at its creation after a short timescale and disappear, 
and since they are localized, they do not contribute to the conductivity. 
We call it ``pair-recombination", which is related to the case observed in the 1D random walk problem;  
the number of walkers that take timescale $\tau$ to return for the first time to its origin is known as $\propto \tau^{-\frac{3}{2}}$. 
Therefore, the distribution of localized NIDWs and those of longer lifetimes differ. 
At the same time, without considering this pair recombination, 
we can naturally expect that there is a characteristic lifetime of NIDWs $\tau_{\rm dw}$ that may depend on temperature or model parameters, 
and the number of NIDWs will follow $\propto e^{-t/\tau_{\rm dw}}$. 
By combining these two factors, we expect
\begin{equation}\label{eq:ndw-fit}
N_{\rm dw}(t) \propto ({\rm const.}+t^{-\frac{3}{2}})e^{-\frac{t}{\tau_{\rm dw}}}. 
\end{equation}
The inset of Fig.~\ref{analysis}(a) shows that $N_{\rm dw}$ is fitted very well with Eq.(\ref{eq:ndw-fit}).
Because the second term from the pair-recombination will be suppressed at a long enough timescale in integrating 
$M_{\rm dw}(t) \propto \int_t^{\infty} \left({\rm const.}+t^{-\frac{3}{2}}\right)e^{-\frac{t}{\tau_{\rm dw}}}dt$, we expect 
\begin{equation}\label{eq:mdw-exp}
M_{\rm dw} (t) \simeq N_{\rm dw}^{\rm tot} e^{-t/\tau_{\rm dw}},\quad   (t \gtrsim \tau_{\rm dw} ). 
\end{equation}
In Fig.~\ref{analysis}(a), we plot $M_{\rm dw}(t)$ obtained by our MC calculation for $N=1024$, $\Delta=1.17$ and $T=0.5$, 
near the boundary of the NI transition. 
We find that Eq.(\ref{eq:mdw-exp}) gives a good approximation, 
which gives $\tau_{\rm dw}=2.19\times 10^3$ and $N_{\rm dw}^{\rm tot}=3.39\times 10^6$. 
The physical implication of these parameters is the typical lifetime and the total number of NIDWs created 
during the entire MCSs, which safely avoided the pair-recombination. 
We notice that within our calculation, the maximum displacement of NIDWs is about less than 100 in a unit of lattice constant. 
We have prepared the system size of $N=512,1024$, which is much larger than these displacements. 
\par
Previously, we have introduced $n_{\rm dw}$ in Eq.(\ref{eq:number_density}) by using the 
values $\tau_{ave}$ and $N_{ave}^{\rm tot}$ averaged over all NIDWs. 
However, we want to extract the NIDWs that contribute to the conduction, and 
accordingly, we adopt the following formula which excludes the effect of pair recombination as 
\begin{equation}\label{eq:number_density2}
  n_{\rm dw} = N_{\rm dw}^{\rm tot} \frac{\tau_{\rm dw}}{N_{\rm MCS}} \frac{1}{N}. 
\end{equation}
\par
Figure~\ref{analysis}(b) shows $\langle x_t^2\rangle$ as a function of $t$.
As mentioned earlier, as long as NIDW follows the normal diffusive mechanism, 
$\langle x_t^2\rangle$ becomes proportional to $t$. 
Indeed, when $t\gtrsim 10^4$, it extrapolates to a linear line of $t$ as 
$\langle x_t^2\rangle \sim 0.07t$ and the diffusion constant $D$ can be extracted as $D=3.5\times10^{-2}$. 
\begin{figure}[htbp]
  \includegraphics[width=7cm]{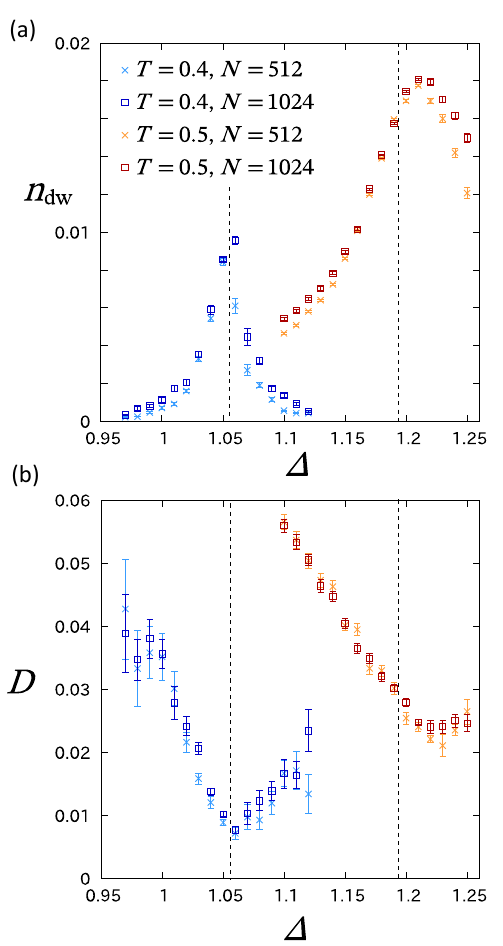}
  \caption{Number density of NIDWs $n_{\rm dw}$ and diffusion constant $D$ obtained by a set of analysis given in Fig.~\ref{analysis} 
at $T=0.4,0.5$ and $N=512,1024$ as functions of $\Delta$.
The dotted line shows the NI transition point $\Delta_c$ for each temperature calculated by Eq.(\ref{eq:boundary}), 
derived by assuming the homogeneous states of I- and N- phases.}
\label{method2}
\end{figure}
\subsubsection{Diffusion constant and number density}\label{sec:results_nD}
Figure~\ref{method2}(a) shows the $\Delta$ dependence of 
$n_{\rm dw}$ obtained from Eq.(\ref{eq:number_density2}) for $T=0.4, 0.5$ and $N=512,1024$. 
At $T=0.4$, $n_{\rm dw}$ shows a sharp peak at around $\Delta_c$, indicating that the 
the number of NIDWs increases significantly toward the transition from both sides, 
signaling the first-order transition between the N and I phases. 
At $T=0.5$, the emergent NIDW phase mentioned in Sec.\ref{subsec:phase_diagram} 
suppresses these peak structures and the number of NIDWs increases, 
taking a maximum at $\Delta_{\text{m}} \simeq 1.22$. 
As we see from the phase diagram in Fig.~\ref{phasediag}(a), $\Delta_{\text{m}}$ 
falls on the NIDW-to-N phase boundary, and is off the ideal NI phase boundary line, 
$\Delta_c=1.19$, shown in broken line. 
\par
Figure~\ref{method2}(b) shows the corresponding values of $D$ 
obtained from $\langle x_{t}^{2} \rangle$ using Eq.(\ref{eq:D}). 
We find a suppression of $D$ at around $\Delta_c$. 
The value of $D$ is larger for the I phase than the N phase 
and takes the minimum at $\sim \Delta_{\text{m}}$, i.e. the same point as the maximum of $n_{\rm dw}$. 
\begin{figure}[tbp]
  \includegraphics[width=8cm]{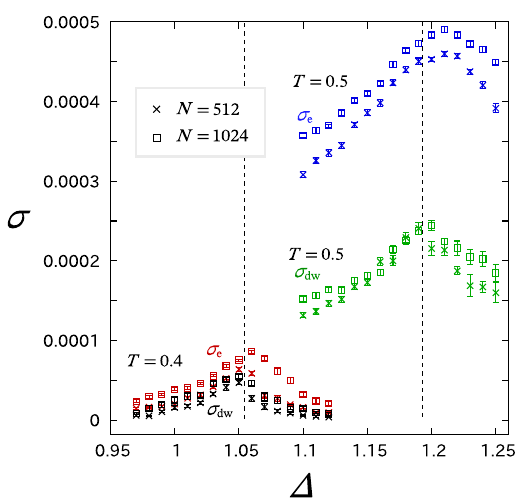}
  \caption{Conductivity $\sigma_e$ and $\sigma_{\rm dw}$ derived from two different frameworks 
at $T=0.4,0.5$ and $N=512,1024$ as functions of $\Delta$.
The dotted line shows the NI transition point $\Delta_c$ for each temperature calculated by Eq.(\ref{eq:boundary}), 
derived by assuming the homogeneous states of I- and N- phases.}
\label{sigma}
\end{figure}
\subsection{Electrical conductivity}
Finally, we derive the conductivity calculated by the two MC schemes in Sec.\ref{MC_dynamics}. 
Figure~\ref{sigma} shows $\sigma_{e}$ and $\sigma_{\rm dw}$ 
for $T=0.4, 0.5$ for the two system sizes. 
Here, $\sigma_e$ obtained from Eq.(\ref{eq:sigmae}) has roughly twice as large amplitudes as $\sigma_{\rm dw}$ 
evaluated via Eq.(\ref{eq:sigmadw}) based on Einstein relation, 
while except for this factor-2, they agree with each other. 
All these results show a clear enhancement of $\sigma_{e/\text{dw}}$ at around the phase boundary, 
successfully reproducing the experimentally predicted enhancement of electrical conductivity. 
The maximum of $\sigma_{e/\text{dw}}$ takes place at 
$\Delta_e$ which is very close to the maximum/minimum position $\Delta_{\text{m}}$ of $n_{\text{dw}}/D$. 
\par
\begin{figure}[tbp]
  \includegraphics[width=8.5cm]{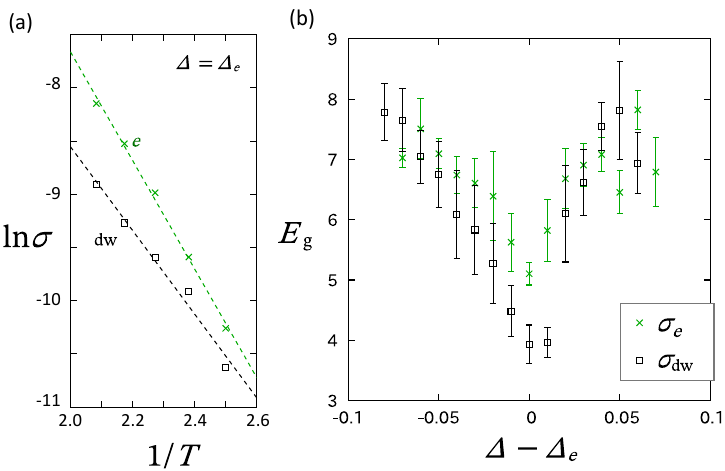}
  \caption{Energy gaps of NIDW for our two calculation schemes are plotted to modified ionic potential, 
$\Delta - \Delta_{\text{m}}$ ($\Delta_{\text{m}}$ is the location where $\sigma$ takes the maximum and 
$\Delta_{\text{e}}\sim\Delta_{\text{m}}$), 
calculated by using Arrhenius plots of the electric conductivities with referring to \cite{Takehara2019}}
\label{arrhenius}
\end{figure}
\subsection{Energy gap of NIDW}
We finally evaluate the activation gap of the conductivity from the obtained conductivities. 
We adopt the Arrhenius plot of $\sigma_e$ and $\sigma_{\text{dw}}$ for $T=0.40-0.48$ at $N=256$, 
as shown in Fig.~\ref{arrhenius}(a). 
The activation gap obtained as in Fig.~\ref{arrhenius}(b) 
shows very rapidly suppressed toward $\Delta_{e}$ from both sides. 
The minimum value of the gap is $E_g=5.1\pm0.2$ for $\sigma_e$ and $E_g=3.9\pm0.3$ for $\sigma_{\rm dw}$. 
The behavior qualitatively agrees with the soliton excitation energy obtained by the bozonization approach\cite{Tsuchiizu2016}. 
\section{Summary and discussion}\label{conclusion}
We analyzed the dynamics of the classical one-dimensional model featuring neutral-ionic(NI) transition 
based on the Monte Carlo(MC) simulation 
to clarify the origin of the enhancement of electrical conductivity observed in the experiments of TTF-CA. 
Our model corresponds to the strong coupling limit ($t_r/U\rightarrow 0$) of the ionic extended Hubbard model. 
We focus on the degree of freedom called neutral ionic domain wall (NIDW), 
which is the bonds that have average $\pm e/2$ charges on two adjacent sites measured from the neutral configuration. 
Although naively the decrease of $\Delta/(U-2V)$ will drive the N-to-I transition as understood from the previous studies, 
the finite temperature phase diagram shows another thermodynamically stable phase, the NIDW phase, 
the mixture of local N and I states separated by the domain walls, 
in between the N and I phases when the temperature is higher than the triple point. 
At this point, the NI phase boundary branches to the two N-to-NIDW and NIDW-to-I phase boundaries 
which are of first order up to the endpoint where the jump in the degree of 
charge transfer $\rho$ vanishes; 
this $\rho$ was determined as the global minima of the free energy obtained by the histogram MC calculation. 
\par
Based on this phase diagram, we performed the MC calculation in the vicinity of the phase boundaries 
to analyze the dynamics of NIDWs by applying two schemes, 
the Kawasaki-like dynamics that apply the electric field as gradients of the site potentials, tracking the motion of particles, 
and the Glauber dynamics in a zero field tracking the NIDWs from their pair creation to the annihilation. 
The $\Delta$ dependence of the two electric conductivities, $\sigma_e$ and $\sigma_{\text{dw}}$, 
obtained from the two schemes agree well. 
In particular, the one at a temperature slightly higher than the triple point shows an enhancement 
at $\Delta_{\text{m}}$ that approximately falls on the NIDW-to-N phase boundary. 
Their values are about five times larger than the ones at the NI phase boundary at low temperatures, 
which cannot be simply explained 
by the $\sigma_{\text{dw}}\propto T^{-1}$ dependence expected for metals. 
\par 
Let us expand the discussion on the nature of activated NIDW conductivity in more detail. 
In our results, the conductivity $\sigma_e$ of charge $e$ of the particles driven by the electric field 
gives the same model-parameter dependence as the conductivity $\sigma_{\text{dw}}$ of the NIDWs carrying $\pm e/2$ charges. 
Therefore, it is natural to understand the origin of this conduction as the diffusive motion of NIDWs. 
Here, $\sigma_{\text{dw}}$ is the product of the number of active domain walls $n_{\text{dw}}$ 
and the diffusion constant $D$, where we find that $n_{\text{dw}}/D$ increases/decreases at 
$\Delta_e\sim \Delta_{\text{m}}$; the increase of $n_{\text{dw}}$ is dominant and enhances $\sigma_{\text{dw}}$. 
At the same time, the activation gap evaluated from the conductivity shows a significant decrease 
at $\Delta_{\text{m}}$. 
The physical implication of these results is that 
the number of NIDWs is activated by the suppression of the excitation gap, 
while the mobility $\mu$ of NIDW or equivalently the diffusion constant $D$ is suppressed. 
Because the mean distance between NIDWs is too large to interact with each other even at $\Delta\sim \Delta_{\text{m}}$, 
it is unlikely that the suppression of mobility is caused by the interactions between NIDWs. 
However, the true object that moves around is the particle, which suffers a competition between N and I configuration 
particularly near the NIDW, which makes the NIDWs less mobile. 
\par
The suppression of $\mu$ or $D$ is more distinct when the system is closer to 
the N phase than the I phase, 
because $\Delta_{\text{m}}$ is at the NIDW-to-N boundary. 
One possible cause could be the increasing number of immobile structures interrupting 
the movements of NIDWs, which is the ``parallel-spin I-domain''; 
the two adjacent D and A sites are occupied by a single particle with the same spin orientation, 
which cannot hop to each other due to the Pauli blocking effect. 
The bottom panel of Fig.~\ref{NIDW}(d) show one of such configurations. 
We indeed observe the duration of dynamics and find that the parallel-spin I-domains interrupt the movements of NIDWs. 
Because the N phase is nonmagnetic, it is much easier to create such a structure stochastically 
at the repetitive creation and annihilation of NIDWs. 
\par
We now discuss the relevance to the experimental findings. 
The temperature-pressure phase diagram of the TTF-CA consists of three phases, N phase, I$_{\text{para}}$ phase and 
I$_{\text{ferro}}$ phase. 
At ambient pressure, the N phase at high temperature undergoes a direct first-order 
phase transition to the I$_{\text{ferro}}$ phase observed by the kink in the resistivity. 
The NQR shows the shift and split of peaks signalling the charge transfer as well as the dimerization, 
and there, the ferroelectricity can happen when the inversion symmetry is broken by the lattice distortion 
which makes the charge transfer more rigid. 
When increasing pressure, the high-temperature N phase crossovers to the I$_{\text{para}}$ phase, 
because the value of $\Delta$ is considered to be effectively suppressed. 
The I$_{\text{para}}$ phase is basically a Mott insulator at high temperature 
which is paramagnetic, and the ferroelectricity is absent. 
The lowering of temperature drives the spin-Peierls transition and 
the system transform to the dimerized I$_{\text{ferro}}$ phase. 
The recent experiment reports a strong enhancement of conductivity by one order of magnitude 
from the one in the I$_{\text{para}}$ phase at around the N-I$_{\text{para}}$ crossover region\cite{Takehara2019}. 
In this region, the infrared spectroscopy shows the active $a_g$ mode related to the dimerization\cite{Okamoto1989,Masino2007} 
which was interpreted as dimer liquid, a fluctuating dimerization. 
The detailed analysis of the temperature dependence of conductivity along the NI crossover line 
shows an activating behavior\cite{Sunami2022}, whose activation energy is $\sim  0.055$ eV, 
one order of magnitude smaller than the charge-transfer excitation energy $0.6-0.7$ eV. 
The average $n_{\text{dw}}$ distribution is derived as one NIDW per 10 sites. 
\par
The present results may apply to the NI crossover region of the TTF-CA phase diagram, $T\sim 280-300$ K and $P\sim 8-10$kbar. 
The energy scale of this temperature is $\sim 0.03$ eV and the activation energy observed from the conductivity measurement is $\gtrsim 0.06$ eV \cite{}. 
In our classical theory, the temperature is $T\sim 0.5$ and the activation energy of NIDW is $E_g\gtrsim 3$ 
in unit of $t_r=0.1$ eV\cite{Ishibashi2010}. 
Previous QMC study including the quantum hopping term $t_r=1$ shows that 
the charge and spin gaps for the same values of $U=7.5$ and $V=3.5$ is $E_g\sim 2$ ($\sim 0.2$ eV) at $\Delta\sim 1$ , 
and does not vanish when $\Delta$ is varied in the range where our phase diagram crosses the boundaries\cite{Nagaosa1986-2}. 
Notice that it is not too different from the first principle evaluation of a gap, $\sim 0.15$ eV\cite{Ishibashi2010}. 
Compared to our case, the transfer integral $t_r$ typically reduces a gap but it is still much larger than the temperature range of our target. Therefore, the activating behavior is guaranteed and we can safely expect the picture we showed in Fig.~\ref{cartoon}(b). 
Experimentally, it has been discussed that the lattice dimerization that allows for another domain wall 
of spins, called spin soliton, are intrinsic to the conductivity as 
they emerge at the relatively high density of one soliton per 20-50 sites. 
This was explained as such that the NIDWs created in pairs from the I phase are recombined 
by the electric current rather than separate further\cite{Sunami2022}, 
which is rather based on the ballistic picture of the motion of particles, not on the diffusive one. 
However, our calculation shows that the stochastic process of creating the NIDWs 
are not as simple, 
and with the aid of their thermal fluctuation and correlation, 
and apart from the NIDWs that disappear by a pair-recombination at a relatively short lifetime, 
there are substantial portion of $n_{\text{dw}}$ that contribute the diffusive conductivity. 
It is also notable that the $\sigma_{e}$ is basically larger than $\sigma_{\text{dw}}$ 
showing that the application of an electric field does not particularly 
support the pair-recombination, contrary to the previous intuitive claims. 
The present results indicate that the activated conductivity can solely be explained 
without the dimerization effect. 
Still, the effect of dimerization on the polarization and conductivity is an issue of wide 
interest not only in NI systems \cite{Egami1993,Resta1995}. 
Our model can be extended further to include the effect of lattice distortion 
which would be our future perspective. 
\par
We finally revisit the diffusive long-timescale behavior of NIDWs, $\langle x_t^2\rangle \propto t$, 
that naturally yields the dynamical exponent, $z=2$. 
As mentioned in Sec.\ref{Sec:dynamics} our conductivity relies on the local fluctuation of particles following Fick's law, 
while neglecting the contributions from the heat transport and the couplings of particles with heat carriers. 
A more simplified description is presented in the time-resolved infrared spectroscopy measurements on TTF-CA \cite{Peterseim2015,Dressel2017}, where the dynamics of the photo-induced NIDWs are treated as random walks, 
which explains well how the domain sizes decay with time. 
Such random walk model was studied for trans-polyacetylene and MX chain compounds 
including disorders and traps\cite{Tabata2009}, and yields the diffusive transport by construction. 
Our way of applying Glauber dynamics provides a more detailed material-based framework for the diffusive transport 
that can be applied to wider classes of systems. 
\par
Let us remind that the relevant ionic Hubbard model with $t, U,$ and $\Delta$ is reduced 
in the strong coupling limit to the spin-1 model with assisted spin-exchange (e.g. $S^z_jS^+_jS^-_{j+1}$ terms) 
and the magnetic anisotropy $(S^z_j)^2$ terms\cite{Legeza2006}. 
Our model corresponds to adding the $V$ term as $S^z_jS^z_{j+1}$ types of interactions 
and deleting the quantum fluctuation to this model ($t=0$) to make them an extended three-state Potts model. 
Recently, the nature of dynamics of 1D quantum spin models at infinite-temperature is found to host three classes, 
depending on their long-time decay in the correlation function $\propto t^{-1/z}$\cite{Dupont2020}; 
the diffusive case with dynamical exponent $z=2$, ballistic one with $z=1$, and the superdiffusive $z=3/2$ ones. 
It was shown that most of the $S=1$ isotropic and anisotropic quantum spin models behave diffusive\cite{Dupont2020}, 
in contrast to the $S=1/2$ models with robust anomalous superdiffusive nature\cite{Nardis2021,Roy2023}. 
Although to which class the system belongs depends much on the types of Hamiltonian, 
our $S=1$ related model can fit this scenario, because its quantum version seriously loses the integrability 
and hosts high anisotropy. 
The class of transport may further change depending on 
what kind of terms are added to such effective Hamiltonian, 
e.g. the lattice dimerization that naturally arise in TTF-CA, 
which will be one of the interesting perspectives. 
\section*{Acknowledgements}
This work was supported by JST SPRING, Grant Number JPMJSP2108, 
"The Natural Laws of Extreme Universe"(No. JP21H05191) KAKENHI for Transformative Areas from JSPS of Japan,
 and JSPS KAKENHI Grants No. JP21K03440.

\bibliography{ttfca_ref}
\end{document}